\documentclass[a4paper,11pt]{article}
\usepackage{pos}

\usepackage[symbol]{footmisc}
\usepackage[english]{babel}
\usepackage{graphicx}
\usepackage{graphics}
\usepackage{braket}
\usepackage{bbold}
\usepackage{amsmath}
\usepackage{nicefrac}
\usepackage{dcolumn}
\usepackage{bm}
\usepackage{slashed}
\usepackage{datetime}
\usepackage{mciteplus}
\usepackage{multirow}
\usepackage{siunitx}
\usepackage{booktabs}
\usepackage{color, soul}
\usepackage[usenames,dvipsnames]{xcolor}
\usepackage{float}
\usepackage[utf8]{inputenc}
\usepackage[normalem]{ulem}
\usepackage{mathtools}
\usepackage{setspace}
\usepackage{comment}
\renewcommand{\thefootnote}{\fnsymbol{footnote}}

\newcommand{{\HFNRevo}}{\tt HF-NRevo}

\title{TQ4Q2.0 Fragmentation Functions for Fully Heavy Tetraquark Production}
\ShortTitle{TQ4Q2.0 FFs for Fully Heavy Tetraquark Production}

\author*[a]{Francesco Giovanni Celiberto}

\affiliation[a]{Departamento de Física y Matemáticas, Universidad de Alcalá (UAH), Campus Universitario, \\ Alcalá de Henares, E-28805, Madrid, Spain}

\emailAdd{francesco.celiberto@uah.es}

\abstract{We investigate the fragmentation dynamics underlying the production of fully heavy tetraquarks by means of the TQ4Q2.0 collinear FF set, covering scalar ($0^{++}$), axial-vector ($1^{+-}$), and tensor ($2^{++}$) configurations. 
The fragmentation inputs are determined within NRQCD factorization for the complete set of relevant partonic channels and are subsequently evolved across heavy-flavor thresholds through the HF-NRevo scheme. 
Perturbative uncertainties associated with fragmentation-scale variations are consistently propagated together with nonperturbative effects encoded in color-composite long-distance matrix elements.
Attention is devoted to the axial-vector sector, whose fragmentation pattern makes it especially sensitive to intrinsic-charm contributions at LHC and FCC energies.
{\tt TQ4Q2.0} therefore provides an uncertainty-aware framework for exploring fully heavy tetraquark production over a broad kinematic range and for connecting exotic-hadron spectroscopy with the partonic structure of the proton.}

\FullConference{The 33rd International Workshop on Deep Inelastic Scattering and Related Subjects (DIS2026)\\
4-8 May 2026\\
Bologna, Italy\\}

\begin{document}
\maketitle

\section{Introduction}
\label{sec:introduction}
Heavy-flavor dynamics offers a privileged window into the transition between short-distance QCD interactions and long-distance hadron formation.
The heavy-quark mass provides a natural hard scale, allowing perturbative calculations to be interfaced with nonperturbative information on the formation of physical hadrons.
This separation makes heavy-flavored systems particularly valuable for precision QCD phenomenology, while their potential sensitivity to interactions beyond the Standard Model broadens their relevance to searches for New Physics.
An especially intriguing manifestation of strong-interaction dynamics emerges when the hadronic spectrum is extended beyond ordinary mesons and baryons.
Multiquark candidates, including tetraquark and pentaquark states, provide direct access to unconventional realizations of color binding and raise fundamental questions about how confinement organizes systems with an enlarged valence content.
Fully heavy tetraquarks are particularly appealing in this respect: the presence of heavy constituents introduces perturbatively accessible scales while retaining the genuinely nonperturbative dynamics responsible for bound-state formation.
A quantitative description of heavy-hadron formation nevertheless remains a nontrivial task.
Even in the comparatively well-understood quarkonium sector, often regarded as the ``hydrogen atom’’ of QCD, no single theoretical description accounts for every aspect of hadronization throughout the full kinematic range.
Non-Relativistic QCD (NRQCD)~\cite{Caswell:1985ui,Bodwin:1994jh} provides a systematic factorization framework in which perturbatively calculable Short-Distance Coefficients (SDCs) are combined with nonperturbative Long-Distance Matrix Elements (LDMEs), with contributions organized according to the relevant Fock-state expansion.
Different production mechanisms become relevant across complementary kinematic regimes: heavy-quark-pair production governs the short-distance dynamics at lower transverse momentum, whereas single-parton fragmentation becomes increasingly important toward moderate and large transverse masses~\cite{Cacciari:1994dr}.
The fragmentation description of heavy quarkonia was initially developed through leading-order calculations of the gluon- and heavy-quark-initiated channels~\cite{Braaten:1993rw,Braaten:1993mp} and subsequently refined through NLO corrections~\cite{Zheng:2019gnb,Zheng:2021sdo}.
These ingredients eventually enabled the construction of fragmentation-function (FF) sets consistently formulated within a variable-flavor number scheme (VFNS)~\cite{Celiberto:2022dyf_short,Celiberto:2023fzz}, with analogous developments later applied to $B_c$-meson production~\cite{Celiberto:2022keu,Celiberto:2024omj}.
Their phenomenological applications and comparisons with LHC measurements provided important support for fragmentation-based descriptions at high energies and motivated the development of a more systematic treatment of heavy-flavor evolution.
The same fragmentation strategy has progressively been generalized from conventional heavy hadrons to multiquark states.
The observation of structures in double-$J/\psi$ final states~\cite{LHCb:2020bwg,ATLAS:2023bft,CMS:2023owd} has stimulated interpretations in terms of compact tetraquark configurations~\cite{Zhang:2020hoh,Zhu:2020xni}, in which two heavy-quark pairs produced at short distances subsequently evolve into bound or near-threshold multiquark states.
The derivation of NRQCD initial conditions for gluon fragmentation into fully heavy tetraquarks~\cite{Feng:2020riv} provided a key ingredient for extending VFNS-based fragmentation analyses to the exotic sector.
Along this direction, the {\tt TQHL1.0} family was introduced for heavy-light tetraquarks~\cite{Celiberto:2023rzw,Celiberto:2024mrq}, followed by the {\tt TQ4Q1.x}/{\tt TQ4Q2.0} and {\tt TQHL1.1} releases~\cite{Celiberto:2024mab,Celiberto:2025dfe,Celiberto:2025ziy,Celiberto:2026kks,Celiberto:2024beg}.
These later developments combine all parton channels from NRQCD~\cite{Feng:2020riv,Bai:2024ezn,Bai:2024flh} and broaden the applicability of the framework to fully heavy multiquark configurations.
The resulting methodology has subsequently been applied to an increasingly diverse spectrum of heavy exotic systems, including fully heavy pentaquarks and triply heavy baryons~\cite{Celiberto:2025ipt,Celiberto:2026ooh,Celiberto:2025ogy,Celiberto:2026qiz}.
Complementary analyses of tetraquark production in forward configurations~\cite{Celiberto:2025vra} have further highlighted how the relative importance of gluon- and heavy-quark-initiated fragmentation can expose different features of the proton’s partonic content, with particular sensitivity to possible intrinsic-charm contributions.

Here we investigate collinear fragmentation into fully heavy tetra-quarks by means of the {\tt TQ4Q2.0} FF sets~\cite{Celiberto:2026kks} and the heavy-flavor nonrelativistic evolution (HF-NRevo) ~\cite{Celiberto:2025euy,Celiberto:2024mex_article,Celiberto:2024bxu,Celiberto:2024rxa,Celiberto:2025xvy,Celiberto:2026rzi,Celiberto:2026zss}.
The resulting description combines NLO NRQCD inputs at the initial scale with DGLAP evolution across heavy-flavor thresholds and a systematic estimate of Missing Higher-Order Uncertainties (MHOUs) through a Monte-Carlo-like replica strategy~\cite{Forte:2002fg}.
Together, these ingredients provide an uncertainty-aware framework for studying the fragmentation dynamics of fully heavy multiquark states.

\section{The {\tt TQ4Q2.0} functions}
\label{sec:HFNrevo}

\begin{figure*}[!t]
\centering

   \includegraphics[scale=0.315,clip]{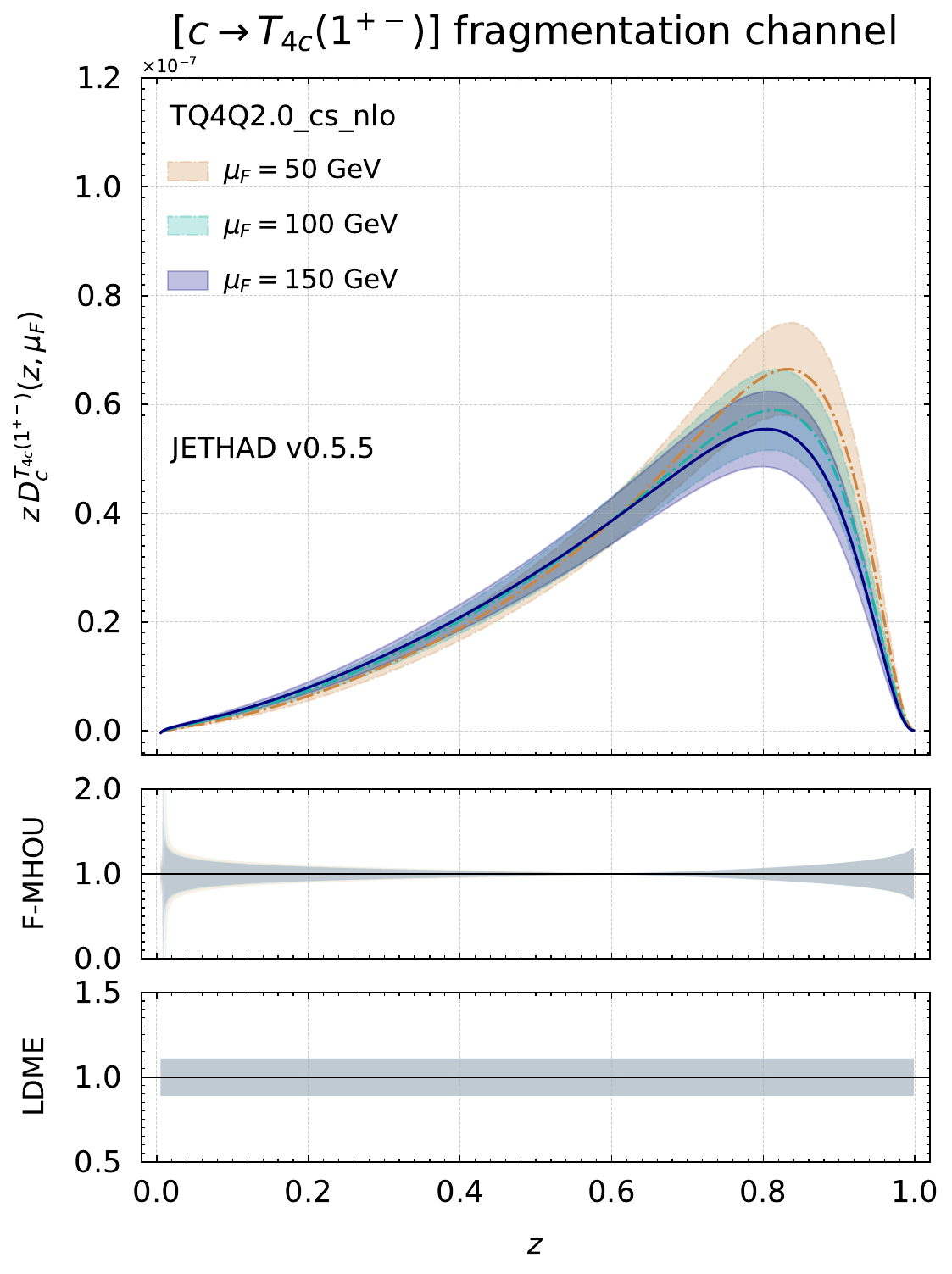}
   \hspace{0.20cm}
   \includegraphics[scale=0.315,clip]{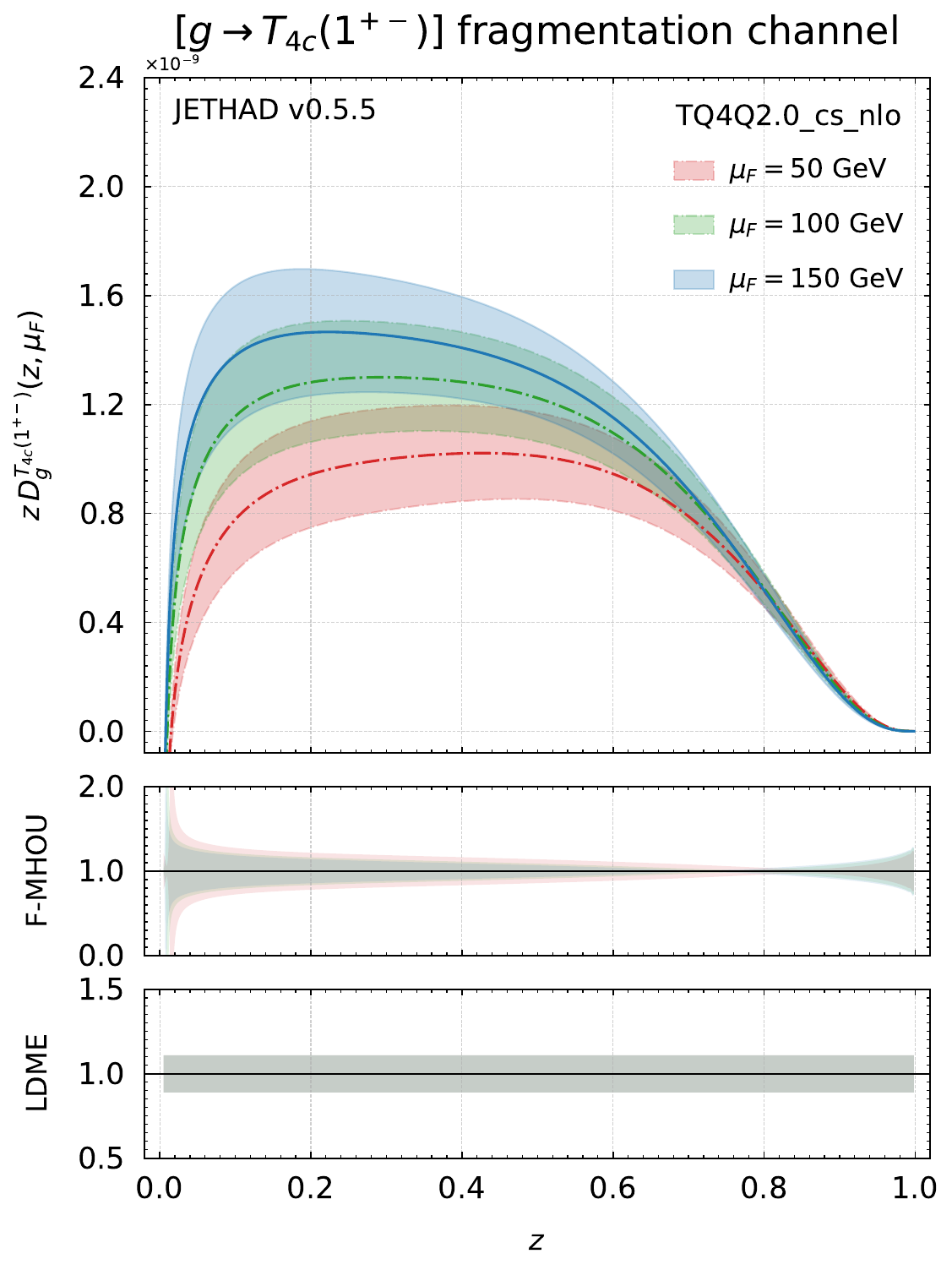}

   \vspace{0.35cm}

   \includegraphics[scale=0.315,clip]{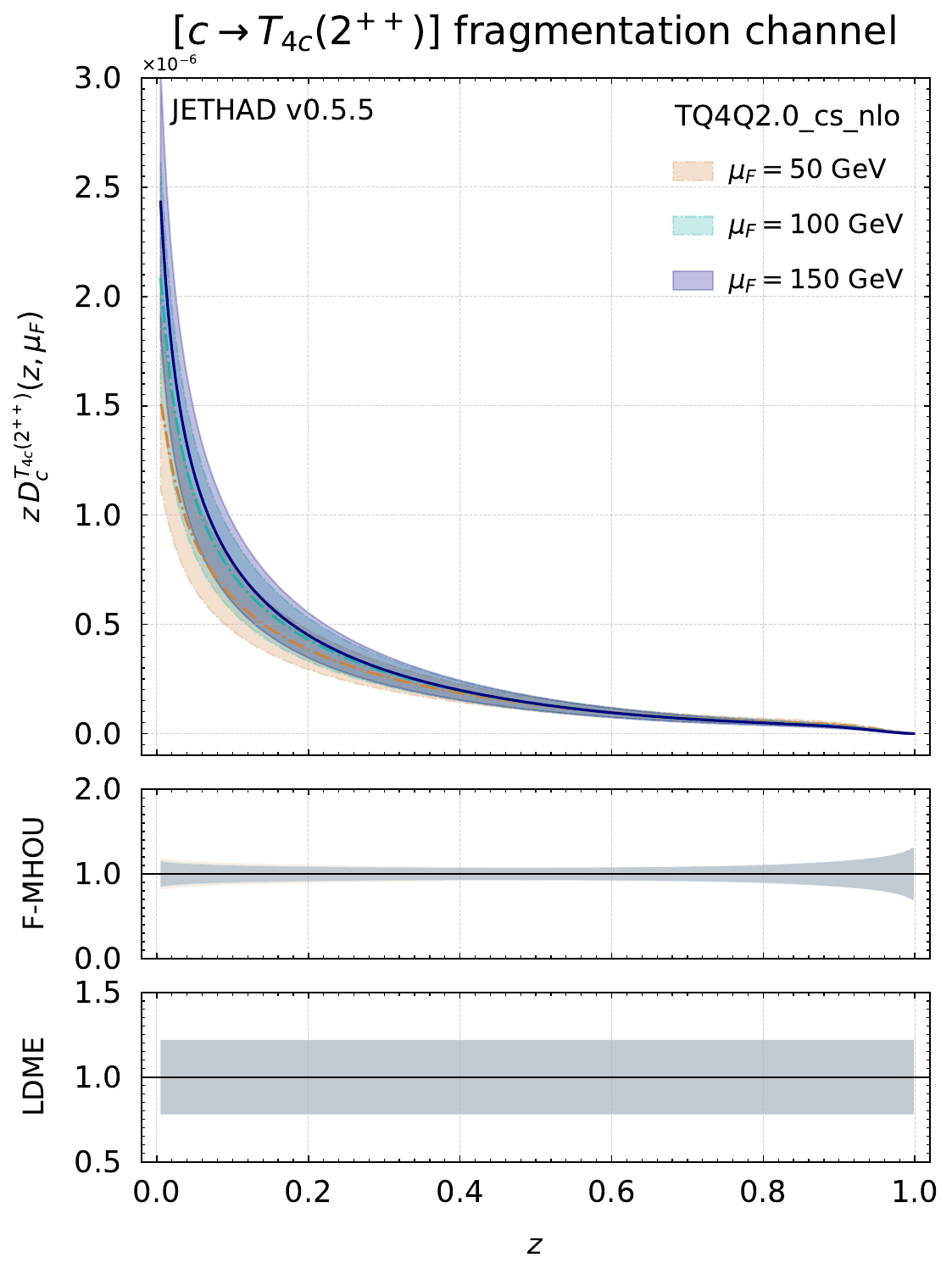}
   \hspace{0.20cm}
   \includegraphics[scale=0.315,clip]{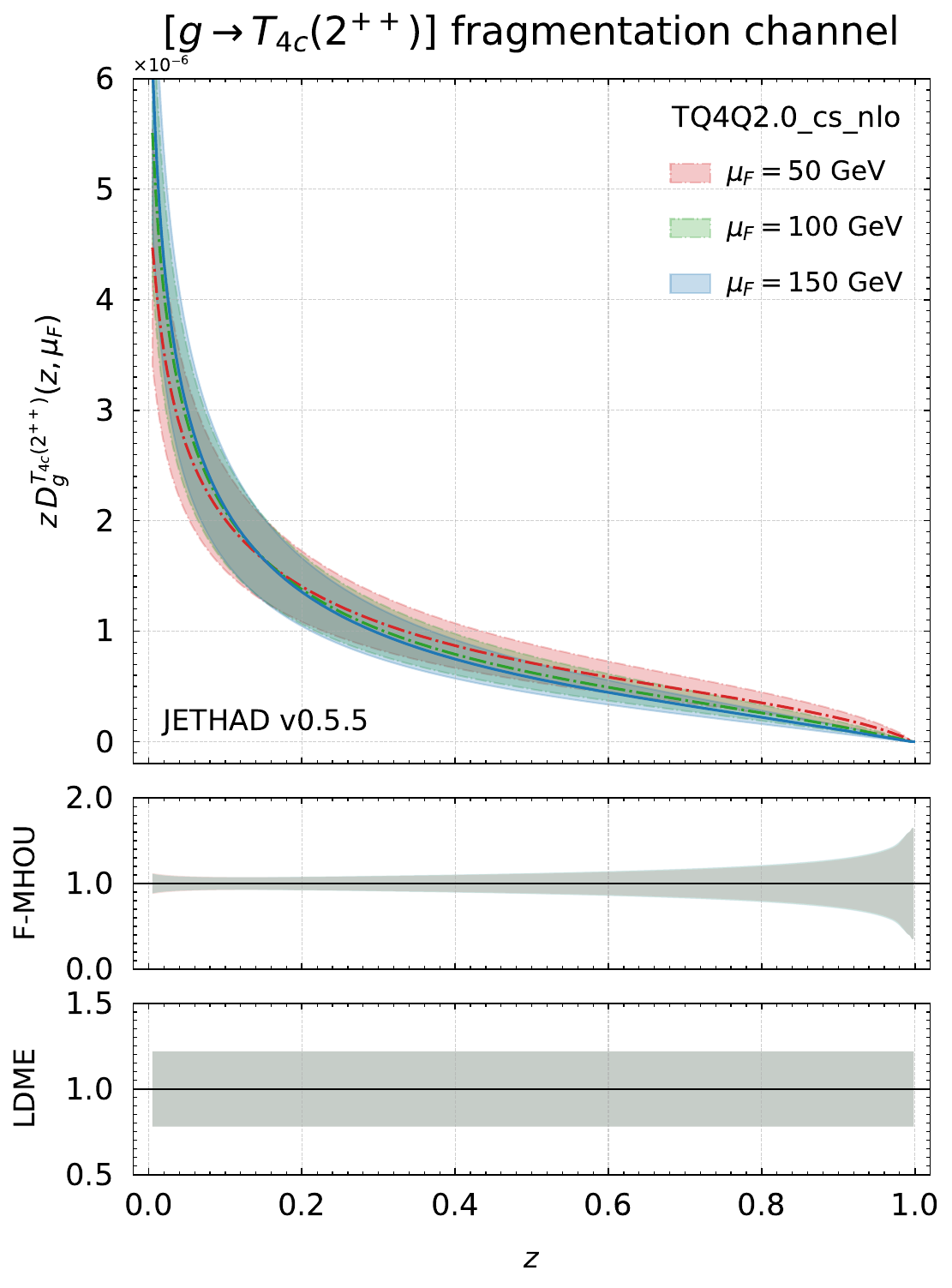}

\caption{{\tt TQ4Q2.0} FFs as functions of $z$ for charm-initiated (left) and gluon-initiated (right) production of axial-vector $T_{4c}(1^{+-})$ (upper) and tensor $T_{4c}(2^{++})$ (lower) states at representative energy scales.
The main panels display the combined F-MHOU and LDME uncertainty bands, while the lower panels disentangle the corresponding F-MHOU replica envelopes and LDME variations, normalized to the respective central predictions.
\vspace{-0.40cm}
}

\label{fig:FFs}
\end{figure*}

The {\tt TQ4Q2.0} family provides a new generation of collinear FFs for fully heavy tetraquarks, built from NRQCD initial-scale inputs and evolved through DGLAP dynamics within a VFNS.
A key improvement from previous releases is the systematic inclusion of the complete set of relevant parton-initiated fragmentation channels at the starting scale.
The corresponding SDCs are combined with color-composite NRQCD LDMEs, thereby separating perturbatively calculable fragmentation dynamics from the nonperturbative information governing tetraquark formation.
A detailed account of the construction of the {\tt TQ4Q2.0} sets, including the analytic inputs and their phenomenological implementation, is given in Ref.~\cite{Celiberto:2026kks}.
The scale dependence of the FFs is determined within the HF-NRevo scheme, specifically designed to evolve nonrelativistic heavy-hadron fragmentation inputs across flavor thresholds.
Its threshold-aware structure permits the different partonic channels to be consistently activated and evolved from their characteristic initial scales, while retaining the physical information encoded in the underlying perturbative fragmentation mechanisms.
The evolution is organized through the complementary symbolic and numerical components of HF-NRevo.
Threshold-sensitive sectors are first treated through the \textsc{symJethad} environment~\cite{Celiberto:2020wpk,Celiberto:2022rfj,Celiberto:2023fzz,Celiberto:2024mrq,Celiberto:2024swu,Celiberto:2026ooh,Celiberto:2025csa,Celiberto:2026zed,Celiberto:2026_TQb_HL_LHC}, after which the complete coupled DGLAP system is numerically evolved with all active partonic species.
Applications of the \textsc{(sym)Jethad} framework to precision QCD phenomenology and hadron structure at low-$x$ can be found in Refs.~\cite{Celiberto:2017ius,Celiberto:2015yba,Celiberto:2016ygs,Celiberto:2022gji,Celiberto:2016hae,Celiberto:2017ptm,Bolognino:2018oth,Celiberto:2020rxb,Celiberto:2022kxx,Celiberto:2020tmb,Celiberto:2023uuk,Celiberto:2023eba,Celiberto:2023nym,Celiberto:2023rqp,Celiberto:2024mdt,Celiberto:2024bfu,Celiberto:2025edg,Bolognino:2021mrc,Celiberto:2021dzy,Celiberto:2021fdp,Celiberto:2022grc} and~\cite{Bolognino:2018rhb,Celiberto:2019slj,Bolognino:2021niq,Celiberto:2018muu}, respectively.
A central ingredient of {\tt TQ4Q2.0} is the systematic treatment of fragmentation missing higher-order uncertainties (F-MHOUs).
Rather than relying on a single scale-variation envelope, perturbative uncertainties are encoded through a replica-like ensemble composed of 100 members.
The ensemble samples the dependence of the initial-scale fragmentation inputs on the relevant perturbative scales, with each member subsequently propagated through the full HF-NRevo evolution.
The resulting distribution of evolved FFs provides an uncertainty band that can be directly propagated to physical observables.
This strategy follows the general philosophy of modern uncertainty treatments based on theory-covariance information~\cite{NNPDF:2024dpb} and Monte Carlo scale variations~\cite{Kassabov:2022orn}; the specific replica construction adopted for {\tt TQ4Q2.0} is detailed in Ref.~\cite{Celiberto:2026kks}.
Perturbative F-MHOUs are complemented by uncertainties associated with the color-composite LDMEs entering the NRQCD initial conditions.
The latter probe the nonperturbative component of tetraquark formation and predominantly affect the normalization of the FFs.
Their combination with the F-MHOU replica ensemble yields the total uncertainty estimate delivered with the {\tt TQ4Q2.0} distributions.
Figure~\ref{fig:FFs} presents representative {\tt TQ4Q2.0} results for charm- (left) and gluon-initiated (right) fragmentation into all-charm tetraquarks at different energy scales.
Upper panels refer to the axial-vector $T_{4c}(1^{+-})$ state, while lower ones focus on the tensor $T_{4c}(2^{++})$ configuration, motivated by the recent CMS spin-parity determination of all-charm tetraquark candidates at the LHC~\cite{CMS:2025fpt}.
Main panels combine F-MHOU and LDME uncertainties, while the ratio panels display their separate contributions relative to the central predictions.
A particularly distinctive feature emerges for the axial-vector $T_{4c}(1^{+-})$ state.
At the initial fragmentation scale, the gluon-initiated NRQCD channel is absent as a consequence of the Landau--Yang constraint, while nonconstituent-quark contributions are strongly suppressed by charge-conjugation properties.
The resulting fragmentation hierarchy therefore enhances the relative importance of the constituent-charm channel, making axial-vector tetraquark production especially sensitive to intrinsic heavy-quark components of the proton wave function~\cite{Ball:2022qks,Guzzi:2022rca,NNPDF:2023tyk}.
This sensitivity is further amplified in forward configurations, where asymmetric partonic kinematics provides enhanced access to the large-$x$ charm content.
As shown in Ref.~\cite{Celiberto:2025vra}, $T_{4c}(1^{+-})$ production can thus serve as a selective probe of intrinsic charm, establishing a direct connection between exotic-hadron production and proton structure and suggesting a complementary interplay between hadron structure and spectroscopy~\cite{Vogt:2024fky}.

\section{Summary and Outlook}
\label{sec:conclusions}
Within the HF-NRevo framework, we constructed the {\tt TQ4Q2.0} collinear FFs for fully heavy tetraquarks from NLO NRQCD inputs across all relevant partonic channels.
Threshold-aware evolution incorporates the multi-scale nature of heavy-flavor fragmentation, connecting initial scales to current and future collider regimes.
Theoretical uncertainties are propagated via a 100-member replica ensemble, combining perturbative F-MHOUs with nonperturbative LDME variations.
The {\tt TQ4Q2.0} release provides an uncertainty-aware baseline for scalar, axial-vector, and tensor configurations.
Its all-parton setup enables detailed studies of channel competition and proton partonic content at the LHC, HL-LHC, and future colliders, with axial-vector states offering a sensitive probe of intrinsic charm.
Broadly, {\tt TQ4Q2.0} advances precision exotic-hadron phenomenology through systematically improvable inputs.
Combining NRQCD factorization, threshold-aware evolution, and robust uncertainty quantification, it provides a solid foundation to explore the production dynamics and internal structure of fully heavy multiquarks.

\section*{Acknowledgments}
\label{sec:acknowledgments}

We are supported by the Atracci\'on de Talento Grant n. 2022-T1/TIC-24176 (Madrid, Spain).

\vspace{-0.05cm}
\begingroup
\setstretch{0.6}
\bibliographystyle{bibstyle}
\bibliography{bibliography}
\endgroup

\end{document}